\documentclass[transmag]{IEEEtran}
\usepackage{latexsym}
\usepackage{graphicx}
\usepackage{amsfonts,amssymb,amsmath}
\usepackage{hyperref}

\usepackage[dvipsnames]{xcolor}
\usepackage{cite}
\usepackage{mathtools}
\usepackage{lipsum}
\usepackage{algorithmic}
\usepackage{textcomp}
\usepackage{booktabs}
\usepackage{esint}

\def\BibTeX{{\rm B\kern-.05em{\sc i\kern-.025em b}\kern-.08em T\kern-.1667em\lower.7ex\hbox{E}\kern-.125emX}}
\begin{document}
\title{Holographic Metasurface Antennas for Uplink Massive MIMO Systems}
\author{Insang Yoo, \IEEEmembership{Student Member, IEEE}, and David R. Smith \IEEEmembership{Senior Member, IEEE}
\thanks{This work was supported by the Air Force Office of Scientific Research (AFOSR, Grant No.: FA9550-18-1-0187) and the National Science Foundation (NSF, Grant No.: 2030068).}
\thanks{The authors are with the Department of Electrical and Computer Engineering, Duke University, Durham, NC, 27708 USA (e-mail: insangyoo1@gmail.com).}}
%\thanks{David~R.~Smith is with the Department of Electrical and Computer Engineering, Duke University, Durham, NC, 27708 USA.}}

\IEEEtitleabstractindextext{\begin{abstract}We propose an uplink massive MIMO system using an array of holographic metasurfaces as a sector antenna. The antenna consists of a set of rectangular waveguide-fed metasurfaces combined along the elevation direction into a planar aperture, each with subwavelength-sized metamaterial elements as radiators. The metamaterial radiators are designed such that the waveguide-fed metasurface implements a holographic solution for the guided (or reference) mode, generating a fan beam towards a prescribed direction, thereby forming a multibeam antenna system. We demonstrate that a narrowband uplink massive MIMO system using the metasurfaces can achieve the sum capacity close to that offered by the Rayleigh channel at 3.5 GHz. We show that metasurfaces supporting multiple fan beams can achieve high spatial resolution in the azimuth directions in sub-6 GHz channels, and thereby form uncorrelated MIMO channels between the base station and users. Also, the proposed metasurface antenna is structurally simple, low-cost, and efficient, and thus is suitable to alleviate RF hardware issues common to massive MIMO systems equipped with a large antenna system.\end{abstract}

}

\maketitle

% \begin{abstract}
% We propose an uplink massive MIMO system using an array of holographic metasurfaces as a sector antenna. The antenna consists of a set of rectangular waveguide-fed metasurfaces combined along the elevation direction into a planar aperture, each with subwavelength-sized metamaterial elements as radiators. The metamaterial radiators are designed such that the waveguide-fed metasurface implements a holographic solution for the guided (or reference) mode, generating a fan beam towards a prescribed direction, thereby forming a multibeam antenna system. We demonstrate that a narrowband uplink massive MIMO system using the metasurfaces can achieve the sum capacity close to that offered by the Rayleigh channel at 3.5 GHz. We show that metasurfaces supporting multiple fan beams can achieve high spatial resolution in the azimuth directions in sub-6 GHz channels, and thereby form uncorrelated MIMO channels between the base station and users. Also, the proposed metasurface antenna is structurally simple, low-cost, and efficient, and thus is suitable to alleviate RF hardware issues common to massive MIMO systems equipped with a large antenna system.
% \end{abstract}

% \begin{IEEEkeywords}
% 2D/3D Metamaterial Intelligence, Information Metamaterials, Aperture Antennas.
% \end{IEEEkeywords}

\section{Introduction} \label{sec:introduction}

\IEEEPARstart{R}{}ecently, multiple-input, multiple-output (MIMO) systems with a large number of base station antennas, referred to as massive MIMO systems, have gained considerable interest due to their capabilities to meet growing demands for high data rates \cite{marzetta2010noncooperative,larsson2014massive,lu2014overview}. In massive MIMO systems, the base stations are equipped with a large number of antennas\textemdash in contrast to conventional point-to-point MIMO or multi-user MIMO systems with fewer than 10 antennas\textemdash to exploit uncorrelated and even asymptotically orthogonal subchannels that can be provided by the large antenna system \cite{bjornson2019massive}. Due to a significant improvement in the spectral efficiency predicted by theoretical studies and prototypes \cite{marzetta2010noncooperative,larsson2014massive,lu2014overview,vieira2014flexible}, massive MIMO systems are considered key ingredients of current and future wireless communication systems.

The large number of radiating elements in massive MIMO systems, however, poses challenges such as the increased cost, design complexity, and high power consumption associated with array antenna systems and the large number of RF chains (i.e., one for each radiator) \cite{bjornson2019massive}. Such challenges become even more significant as the number of antennas required in the MIMO system increases. Accordingly, there have been approaches to alleviate the challenges in array antenna and RF hardware implementation; for instance, a hybrid beamforming architecture for millimeter-wave massive MIMO systems has been proposed to reduce the number of required RF chains by exploiting the sparsity of millimeter MIMO channels \cite{sun2014mimo}. Also, the antenna selection technique has been studied to reduce hardware complexity and improve efficiency by considering the unequal contribution of radiating elements to MIMO channels \cite{gao2015massive}. In most sub-6 GHz bands, a full digital beamformer (DBF) is an attractive solution for most MIMO systems as it provides the greatest flexibility in creating and adapting multiple beams to the propagation channels \cite{bjornson2019massive}. However, massive MIMO systems with a large antenna system (e.g., on the order of hundreds) and low power consumption may require alternative choices for scalable and sustainable antenna and RF systems.

Recently, there has been increasing interest in holographic metasurfaces to address the challenges posed by the increased RF complexity of massive MIMO systems \cite{bjornson2019massive}. In holographic metasurfaces, subwavelength-sized scattering (or radiating) elements are distributed over an aperture such that the metasurface can be considered as a continuous distribution of the effective current elements excited by a feed wave. The individual elements are designed using the holographic principles to generate desired scattered fields \cite{glybovski2016metasurfaces,chen2016review}. As the metasurfaces utilize the phase advance of the feed wave and the phase shift imparted by the scattering elements, the metasurface architecture can be passive yet with sufficient degrees of freedom for beamforming and waveform shaping \cite{smith2017analysis}, which is appealing for massive MIMO systems.

Among available holographic metasurface configurations, we find that the waveguide-fed metasurface antennas reported in \cite{smith2017analysis} represent a suitable radiative platform with a simple, low-cost, and low profile configuration that can bring many opportunities to massive MIMO systems. Waveguide-fed metasurface antennas consist of a waveguide that excites metamaterial radiators embedded in its upper conducting layer. Each metamaterial radiator can couple a portion of the energy in the guided mode to a dipole-like radiating mode with desired amplitude and phase at each location, thereby providing the basis for beamforming capabilities \cite{smith2017analysis}. Also, low-power tuning mechanisms such as diodes or liquid crystals can be introduced into the individual metamaterial radiating elements to enable dynamic control over the composite radiation patterns \cite{sleasman2015dynamic,sleasman2017reconfigurable,boyarsky2021electronically}. In this manner, waveguide-fed metasurfaces can avoid using costly and power-hungry RF components and thus have been proposed as a novel radiative platform for various applications \cite{hunt2013metamaterial,gollub2017large,gowda2018focusing,boyarsky2017synthetic,johnson2016extremum,yoo2018enhancing,yoo2020dynamic}.

For massive MIMO systems operating in sub-6 GHz bands, the spatial properties of the propagation channels can be exploited to further reduce the RF system complexity. In particular, it is observed in sub-6 GHz band MIMO channels that the angular spread in elevation is generally small compared to that in the azimuth directions \cite{3gpprel6,asplund2006cost,kyosti2007winner}; therefore, it is natural to consider a holographic waveguide-fed metasurface antenna with high spatial resolution in the azimuth directions. To implement such a metasurface antenna system, we consider an array of rectangular waveguide-fed metasurfaces patterned so as to generate multiple fan beams.

% the directions of arrival and departure
% \cite{{erceg2004tgn},{kyosti2007winner},{correia2001wireless},{asplund2006cost}}
% Inspired by the benefits, we consider a holographic waveguide-fed metasurface antenna as a transmit antenna system for massive MIMO systems.
% low-power driven tuning elements can be 
% By reviewing these methods, 
% Such challenges in hardware implementation can be alleviated by the usage of a hybrid beamforming architecture for milimeter wave massive MIMO systems \cite{sun2014mimo} and full digital beamformer (DBF) for sub-6 GHz massive MIMO systems \cite{bjornson2019massive}.

In this work, we present an uplink massive MIMO system using a set of holographic metasurface antennas to form a base station antenna operating at 3.5 GHz. To demonstrate the operation of the system, we design holographic beamforming metasurface antennas, each consisting of a rectangular waveguide and radiating metamaterial elements, and verify their performance using the coupled dipole method. We then demonstrate that the proposed massive MIMO system using the metasurface aperture generating multiple fan beams can achieve a sum capacity close to that offered by the Raylegh channel.

% We note that in this paper, $\left[\mathbf{A}\right]_{ij}$ represents the element of matrix $\mathbf{A}$ at $i$th row and $j$th column, and $\bar{A}$ is a vector. The operator $(\cdot)^{T}$ and $\text{det}(\cdot)$ are the transpose and determinant operators, respectively.

\section{Design of Holographic Metasurface Antennas} \label{sec:theory}

\begin{figure}[!t]
\centering
\includegraphics[width=3.5in]{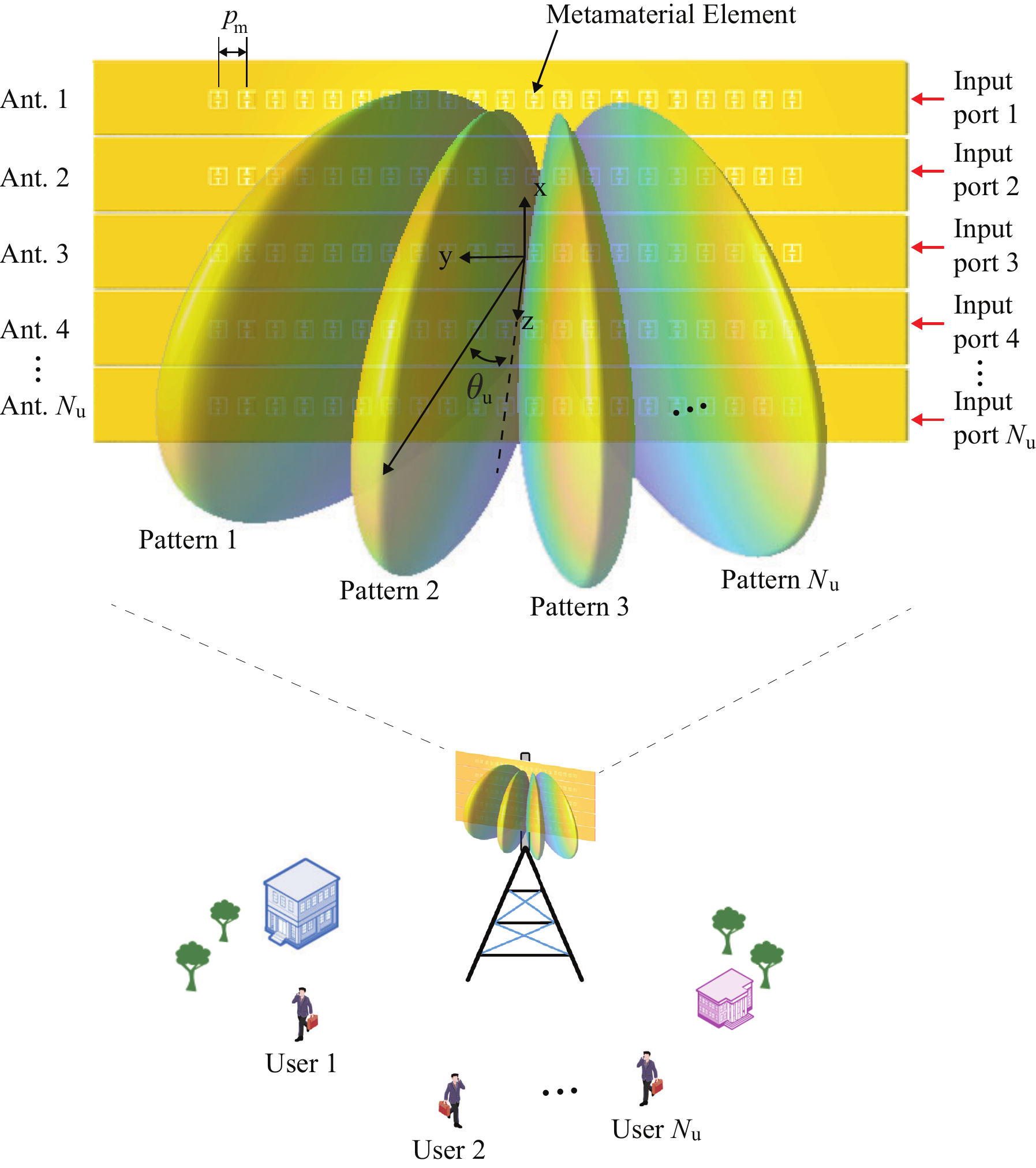}
\caption{Schematic of an array of rectangular waveguide-fed metasurface antenna for uplink massive MIMO system. The antenna consist of $N_{bs}$ metasurfaces with metamaterial elements as radiators, and $u$th metasurface generates a fan beam toward prescribed direction $\theta=\theta_{u}$.}
\label{Fig1_Schematic}
\end{figure}

\subsection{Holographic Beamforming Waveguide-fed Metasurface Antennas} \label{sec:holographic_method}

\begin{figure}[!t]
\centering
\includegraphics[width=2.7in]{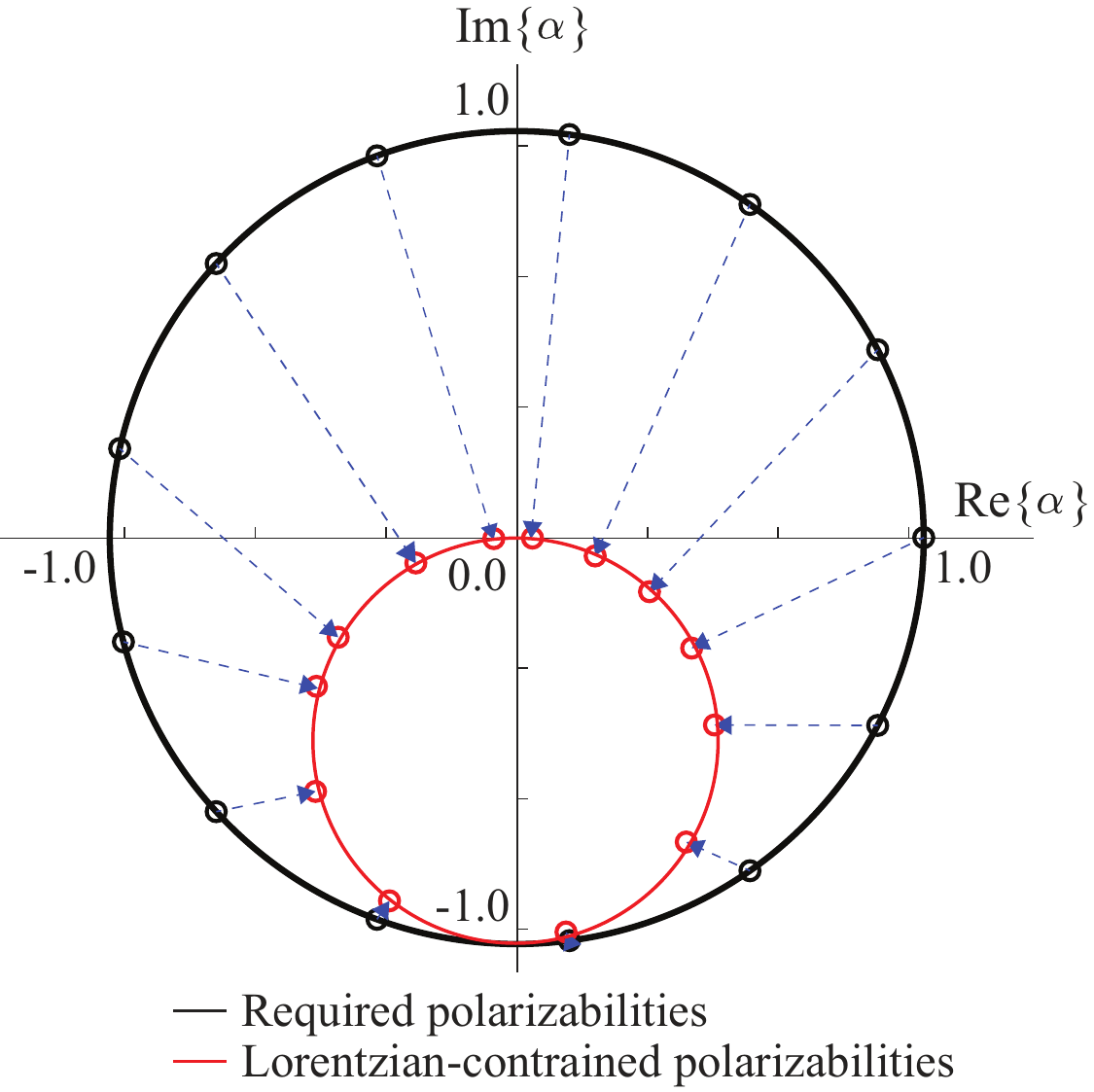}
\caption{Illustration of the Euclidean mapping technique (blue arrows). The required polarizabilities for beamforming (outer, black circles) and the Lorentzian-constrained polarizabilities (red, inner circles) are plotted in the complex plane). The Lorentzian-constrained polarizabilities are found by minimizing the Euclidean distance between the required and accessible polarizabilities.}
\label{Fig2_Euclidean}
\end{figure}

We consider a rectangular waveguide-fed metasurface antenna with a set of metamaterial elements that implement a holographic beamforming solution for the guided modes \cite{smith2017analysis}, as illustrated in Fig. \ref{Fig1_Schematic}. Each metasurface antenna consists of a rectangular waveguide, with resonant metamaterial radiators etched into the upper conductor. In this device, the guided modes excite the metamaterial radiators, forming electric fields on their apertures and leaking a portion of the energy as radiation. Since the metamaterial elements are electrically small, each element forms a subwavelength iris that can be modeled as a combination of polarizable electric and magnetic dipoles \cite{pulido2017polarizability}. If the distance between the elements is also subwavelength, they represent a step-wise approximation to a continuous, linear distribution of current density over a conducting plane. Holographic solutions are then easily identified and implemented as the current density that results from the interference of the incident waveguide mode and the desired radiation pattern back-propagated to the aperture \cite{smith2017analysis,johnson2016extremum,johnson2015sidelobe}.

To facilitate the analysis of the system using the coupled-dipole approach, each metamaterial element is modeled as a polarizable magnetic dipole\textemdash assuming that the electric dipole moment is negligibly small\textemdash and characterized with an effective magnetic polarizability. The required polarizabilities for beamforming are obtained such that the phase advance in the feed waveguide is compensated, and the radiated field constructively interferes at a desired direction \cite{smith2017analysis}. For example, for the metasurface antenna shown in Fig. \ref{Fig1_Schematic}, the desired polarizability of $i$th element located at $y_{i}$ can be expressed as
\begin{equation} \label{ideal_polarizabilities}
\alpha^{m,req}_{x,i}=e^{j \beta y_{i}}e^{jk_{0}y_{i} \cos{\theta_0}},
\end{equation}
where $\beta$ and $k_{0}$ represent the propagation constant and free space wavenumber, respectively. $\theta_{0}$ is the desired beam direction.  However, the available set of polarizabilities supported by resonant metamaterial elements is limited by the Lorentzian resonance \cite{smith2017analysis,pulido2017polarizability}, given by
\begin{equation} \label{lorentzian}
\alpha^{mm}_{xx}\left(\omega\right)=\frac{F_{m}\omega^2}{\omega_0^2-\omega^2+j\omega\gamma_{m}},
\end{equation}
with $\omega$ and $\omega_0$ being, respectively, the angular frequency and the angular resonant frequency. $F_{m}$ and $\gamma_{m}=\omega_0/2Q_m$ represent a coupling factor, damping factor, respectively, and $Q_m$ is the quality factor of the metamaterial element. 

Note that the amplitude and phase of the polarizability in Eq. (\ref{lorentzian}) is inherently coupled. Thus, resonant metamaterial elements cannot implement the required set of polarizabilities in Eq. (\ref{ideal_polarizabilities}). As a result, a simple mapping technique that bridges the gap between the required polarizabilities in Eq. (\ref{ideal_polarizabilities}) and the available polarizabilities have been studied. One useful method is to find the polarizability having the minimum Euclidean distance from the required polarizability, which is often referred to as the Euclidean modulation \cite{boyarsky2021electronically}. A more extensive discussion on polarizability mapping methods can be found in \cite{smith2017analysis}. Fig. \ref{Fig2_Euclidean} illustrates the mapping technique for a beam synthesis, where the required polarizabilities obtained from using Eq. (\ref{ideal_polarizabilities}) are mapped to the accessible, Lorentzian-constrained polarziabilities in Eq. (\ref{lorentzian}). As illustrated in Fig. \ref{Fig2_Euclidean}, the polarizability mapping is achieved by minimizing the Euclidean distance in the complex plane.

% In this work, we employ the Euclidean mapping method for beam synthesis, and its details can be found in \cite{smith2017analysis}.
% , as detailed in \cite{smith2017analysis}.

\subsection{Simulated Holographic Metasurface Antenna using the Coupled Dipole Model} \label{sec:design_part2}

In this subsection, we design metasurface antennas using the coupled dipole method (CDM), an effective analysis and design tool for metasurface antennas \cite{pulido2016discrete,pulido2017discrete,pulido2017polarizability,yoo2019analytic,yoo2020full}. At this point, it should be emphasized that our goal is a conceptual development of an uplink massive MIMO system using holographic beamforming metasurfaces; therefore, we use a generic model of metasurface antennas that are simulated using the CDM and metamaterial elements characterized by Eq. (\ref{lorentzian}). 

We consider here an array of $N_{bs}$ rectangular waveguide-fed metasurfaces stacked along the $\hat{x}$ direction, as shown in Fig. \ref{Fig1_Schematic}. Each metasurface has $N_{m}$ metamaterial elements, each characterized by the Lorentzian parameters $F_{m}=3.0\times10^{-9}$ $\text{m}^3$ and $Q_m=10.0$. The spacing between adjacent elements is $p_{m}=30.0$ mm. We assume that each element can have its resonant frequency $\omega_0/2\pi$ tuned to lie between $3.2$ GHz and $3.8$ GHz. Note that we choose practical values for the Lorentzian parameters that have not been optimized for the simulated antenna design. The substrate filling the waveguide is assumed to be 1.52-mm-thick Rogers 4003C ($\epsilon_{r}=3.55$, $\tan\delta=0.0027$). The width of each rectangular waveguide is $30.0$ mm. The waveguide parameters enable us to analytically determine the waveguide feed mode.

% The extraction of the polarizability of a metamaterial element using full-wave simulations or experiments can be found in \cite{pulido2017polarizability}, a task which is beyond the scope of this paper. 

\begin{figure}[!t]
\centering
\includegraphics[width=3.3in]{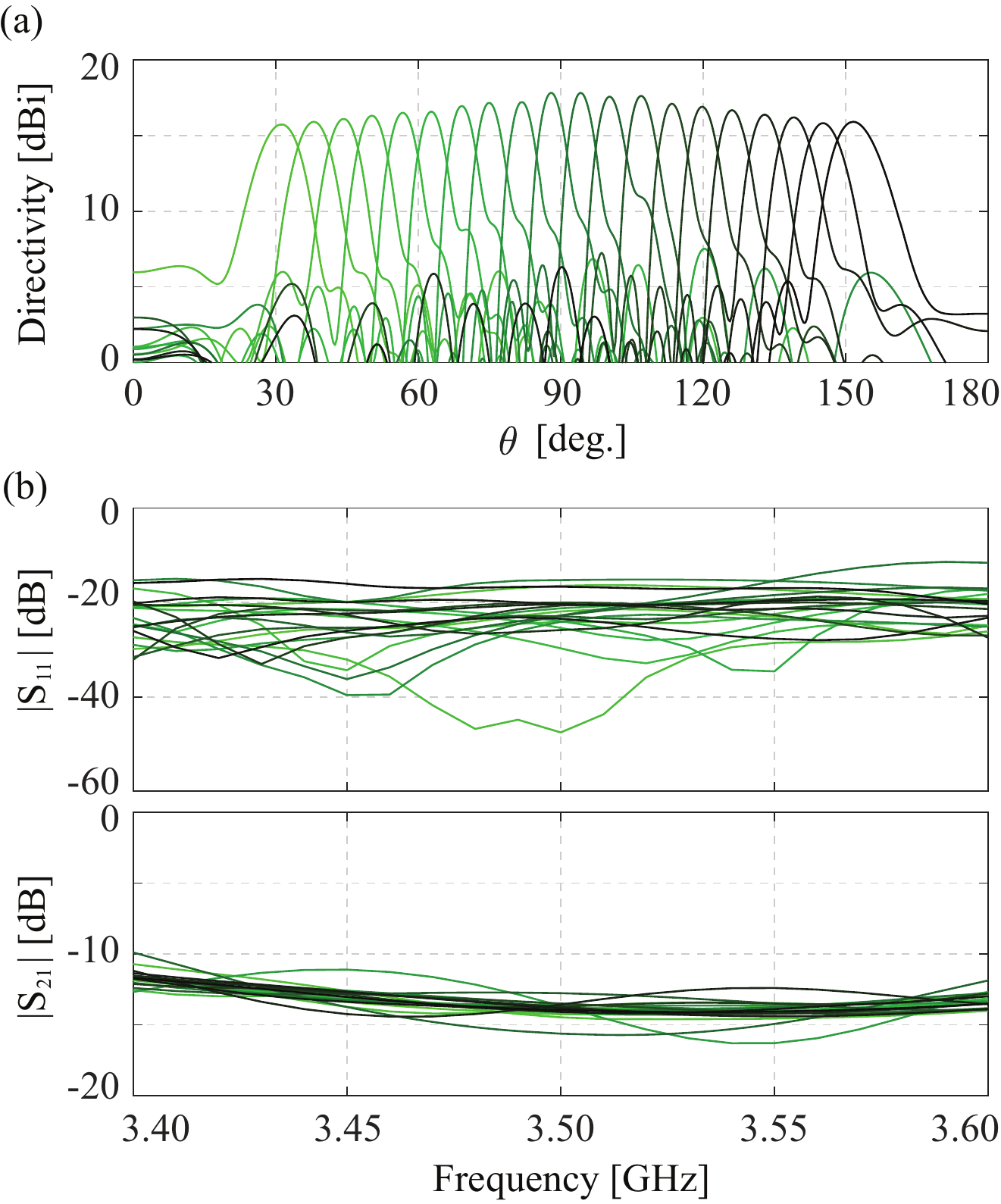}
\caption{(a) Directivity patterns of each  metasurface antenna at the target frequency of $3.5$ GHz. (b) S-parameters of each antenna.}
\label{Fig3_Patterns}
\end{figure}

With the above parameters selected, the CDM allows self-consisting modeling of the metasurface using the following matrix equation, given in \cite{bowen2012using,pulido2017polarizability} as
\begin{equation} \label{cdm_eq}
\mathbf{G}^{mm}_{xx} \mathbf{m}_{x}=\mathbf{H}^{inc}_{},
% \left[\left(\alpha^{m}_{x,i}\right)^{-1}\delta_{ij}-\mathbf{G}^{mm}_{ij}\right] \mathbf{m}_{x}=\mathbf{H}^{inc}_{},
\end{equation}
where $\mathbf{m}_{x}, \mathbf{H}^{inc} \in\mathbb{C}^{N_{m} \times 1}$ are the effective magnetic dipole moments representing the metamaterial elements and the incident magnetic fields, respectively. The off-diagonal entries of the matrix $\mathbf{G}^{mm}_{xx} \in\mathbb{C}^{N_{m} \times N_{m}}$ are the Green's functions \cite{collin1960field}, and the diagonal entries are the inverse of the polarizabilities, i.e., $\left(\alpha^{m}_{x,i}\right)^{-1}$. The matrix equation takes into account all dipolar interactions between the elements and the waveguide in a metasurface antenna. Note that we apply the Lorentizan-constrained polarizabilities in Eq. (\ref{cdm_eq}), which are obtained from the Euclidean mapping method of the required polarizabilities in Eq. (\ref{ideal_polarizabilities}) to verify the beamforming capability of the designed metasurfaces. By solving the matrix equation in Eq. (\ref{cdm_eq}) for the magnetic dipole moments, we can compute the radiation pattern and S-parameter using the magnetic dipole's Green's functions \cite{pulido2017discrete}.

Fig. \ref{Fig3_Patterns}(a) shows the simulated directivity patterns (i.e., radiation patterns on the $yz$ plane) of the designed antenna with $N_{bs}=20$ metasurfaces, each with $N_{m}=30$ metamaterial elements. As the spacing between adjacent metamaterial elements is $p_{m}=30$ mm, the overall aperture size is $\sim 90$ cm $\times$ $60$ cm. The operating frequency is chosen to be 3.5 GHz, and each metasurface is designed to generate a fan beam toward the prescribed directions, ranging from $\theta=30$ to $150$ degrees (with an interval of $6.32^{\circ}$). As shown in Fig. \ref{Fig3_Patterns}(a), the peak directivities of the patterns are maintained to be larger than $15.74$ dBi for the designed metasurfaces.

Note that each of the radiation patterns shown in Fig. \ref{Fig3_Patterns} is obtained by simulating a metasurface in each row using (\ref{cdm_eq}) and computing its pattern. This approach of simulating an array of metasurfaces ignores the mutual interaction of metamaterial elements\textemdash through the radiated fields\textemdash embedded in waveguides in different rows, which may lead to errors in the predicted radiation patterns. However, such interactions of metamaterial elements are often not significant \cite{pulido2018analytical}. Therefore, we here use the approach to simplify the analysis of the array system rather than introducing an antenna model that accounts for the interactions between all pairs of elements in a metasurface array.

% to generate the radiation patterns of the array system.
% Note that we did not simulate the entire array of metasurfaces to generate the radiation patterns shown in Fig. \ref{Fig3_Patterns}(a). Rather, we simulated the metasurface in each row in Fig. \ref{Fig1_Schematic} by solving (\ref{cdm_eq}) for the dipole moments representing metamaterial elements in the metasurface. This approach of simulating an array of metasurface antennas ignores the mutual coupling of elements\textemdash through the radiated fields\textemdash embedded in different metasurfaces, thus leading to errors in the predicted radiation patterns. However, the mutual interaction of metamaterial elements through the radiated field is generally not significant \cite{pulido2018analytical}. Thus, we here OOO.

Fig. \ref{Fig3_Patterns}(b) shows the S-parameter of the metasurfaces as a function of frequency, where port $``1"$ and $``2"$ of each metasurface indicate the input port (at the right edge of each metasurface) and termination port (at the left edge of each metasurface), respectively. For all of the metasurfaces, it is computed that $S_{11} < -15.62$ dB and $S_{21}<-19.89$ dB at 3.5 GHz. The computed $S_{11}$ indicates that each metasurface has a reasonably good level of impedance match at port 1.

% Also, SLL for each beam is computed and is > 7.78 dB for all beams.

\section{Performance Results in Uplink Massive MIMO Systems} \label{mimo_analysis}

\subsection{Channel Model} \label{sec:channel_model}

We here focus on the analysis of an uplink massive MIMO system where the base station uses the designed metasurface antenna serving one sector in a cell. We assume that adjacent cells do not cooperate. We further assume that each rectangular waveguide-fed metasurface in the antenna system is connected to a single RF chain; thus, the base station has $N_{bs}$ RF chains. The maximum number of active users is $N_{u}$, and each user terminal is equipped with one dipole antenna. We consider a narrowband, flat-fading propagation channel which yields a received signal, expressed as
\begin{equation} \label{mimo_system_eq}
\mathbf{r}=\sqrt{\rho}\mathbf{H}\mathbf{s} +\mathbf{n},
\end{equation}
where $\mathbf{s} \in\mathbb{C}^{N_{u} \times 1}$ is the transmit signal vector normalized such that $\mathbb{E}\left[|\mathbf{s}|^2\right]=1$, and $\mathbf{r} \in\mathbb{C}^{N_{bs} \times 1}$ is the receive signal vector. Also, $\mathbf{n} \in\mathbb{C}^{N_{u} \times 1}$ represents a complex Gaussian noise with zero-mean and unit covariance. $\mathbf{H} \in\mathbb{C}^{N_{bs} \times N_{u}}$ is the channel matrix such that $\mathbb{E}\left[||\mathbf{H}||^2_{F}\right]=N_{bs}N_{u}$ where $||\cdot||^2_{F}$ represents the Frobenius norm. $\rho$ is the signal-to-noise power ratio at the receive antenna.

% $\mathbb{E}\left[\cdot\right]$ indicates the expectation is taken over the transmitted symbols. 

To simulate MIMO propagation channels, we consider a clustered channel model, referred to as the Saleh-Valenzuela model \cite{saleh1987statistical,wallace2002modeling}. Using the channel model, the entry in the $m$th row and $n$th column of the channel matrix $\mathbf{H}$ in Eq. (\ref{mimo_system_eq}) can be written as
\begin{equation} \label{SV_model}
\left[\mathbf{H}\right]_{mn}=\sqrt{\frac{1}{N_{c}N_{r}}}\sum_{p=1}^{N_{c}}\sum_{q=1}^{N_{r}}h_{pq}g^{u}_{n}\left(\theta^{u}_{pq}\right)g^{bs}_{m}\left(\theta^{bs}_{pq}\right),
\end{equation}
where $N_c$ and $N_r$ represent the number of clusters in the propagation environment and rays in each cluster, respectively. $h_{pq}$ is the complex gain of the $q$th ray in $p$th cluster and is assumed to be Rayleigh fading coefficient. $g^{u}_{n}\left(\theta^{u}_{pq}\right)$ and $g^{bs}_{m}\left(\theta^{bs}_{pq}\right)$ are the antenna gain at the $n$th user terminal and $m$th base station, respectively, where $\theta^{u}_{pq}$ and $\theta^{bs}_{pq}$ are the transmit and receive angles of the rays. Note that $\theta^{u}_{pq}$ and $\theta^{bs}_{pq}$ are generated according to a Laplacian distribution with a mean cluster angle $\theta^{u}_{mean}$, $\theta^{bs}_{mean}$ and angular spread $\sigma^{u}_{}$, $\sigma^{bs}_{}$, respectively. In constructing the channel matrix $\mathbf{H}$ using Eq. (\ref{SV_model}), we assume that the clusters are independent, and the mean clusters are generated according to the uniform distribution with $\theta^{u}_{mean}\in\left[0^{\circ},360^{\circ}\right]$ and $\theta^{bs}_{mean}\in\left[-60^{\circ},60^{\circ}\right]$. It should be noted in Eq. (\ref{SV_model}) that we assume that the rays are confined to the $yz$ plane in Fig. \ref{Fig1_Schematic} (i.e., $\phi=90^{\circ}$), as the angle spread in the $xz$ plane is generally small compared to that in the $yz$ plane in sub-6 GHz MIMO channels \cite{erceg2004tgn,kyosti2007winner,spencer2000modeling}.

% For the channel realization, we assume that $\theta^{u}_{mean}\in\left[0^{\circ},360^{\circ}\right]$ and $\theta^{bs}_{mean}\in\left[-60^{\circ}^{\circ},60^{\circ}\right]$, and the mean cluster angles are .

\subsection{Performance Analysis} \label{sec:performance_analysis}

To demonstrate the operation of the proposed uplink massive MIMO system, we simulated the system according to Eqs. (\ref{mimo_system_eq}) and (\ref{SV_model}), where the radiation patterns of the metasurfaces are incorporated to construct the MIMO channel matrix $\mathbf{H}$. To evaluate the performance of the proposed system, we use the condition number $\kappa$ of the matrix $\mathbf{H}$ and the sum capacity over $10,000$ realizations of the channel matrix. Note that $(\cdot)^{H}$ is the Hermitian operator. In constructing the channel matrix, we assume that the number of clusters and rays are $N_{c}=6$ and $N_{r}=11$, respectively, which are used to model macro-cells in urban areas \cite{molisch2004generic}. In the simulations, we further assume that the channel state information is available to the base station, and the users do not cooperate with each other. Under these assumptions, the sum capacity can be expressed as \cite{perez2020analysis,heath2018foundations}
\begin{equation} \label{sum_capacity}
C_{sum}=\sum_{u=1}^{N_{u}} \log_{2}\left(1+\frac{\rho}{N_{u}}\lambda_{u}\right),
\end{equation}
where $\lambda_{u}$ represents the $u$th eigenvalue of the matrix $\mathbf{H}^{H}\mathbf{H}$.

\begin{figure}[!t]
\centering
\includegraphics[width=2.45in]{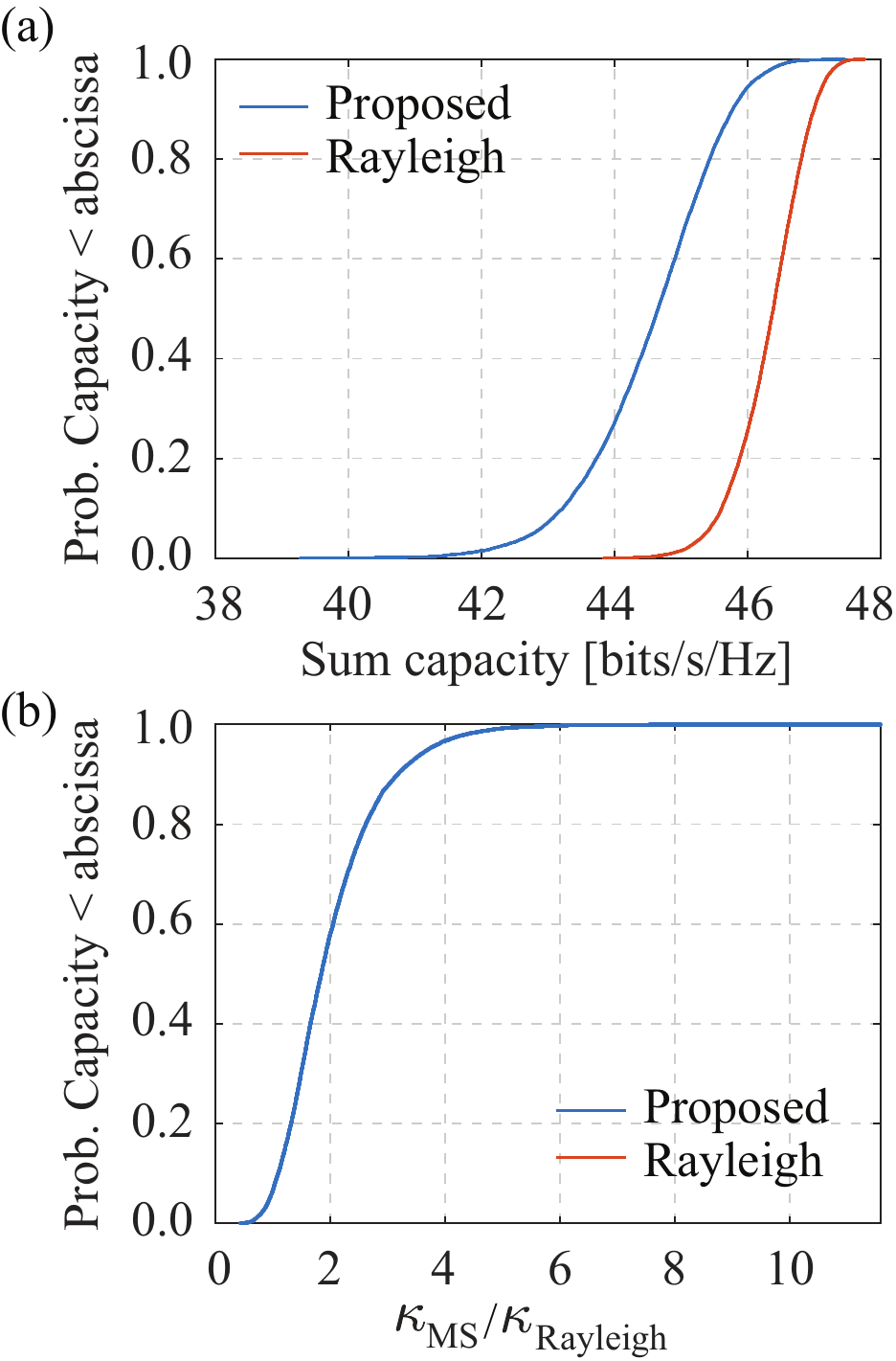}
\caption{The cumulative distribution of (a) the sum capacity and, (b) the ratio of
the condition numbers of MIMO channel matrix, i.e., $\kappa_{MS}/\kappa_{Rayleigh}$. It is assumed that the angular spread is $\sigma^{u}_{}=\sigma^{bs}_{}=20^{\circ}$.}
\label{Fig4_capacity}
\end{figure}

Fig. \ref{Fig4_capacity}(a) shows the cumulative distribution (CDF) of the sum capacity for the angle spread $\sigma^{u}_{}=\sigma^{bs}_{}=20^{\circ}$ and $\rho =10$ dB for the antenna with $N_{bs}=20$ metasurfaces. Each metasurface is assumed to have $N_{m}=30$ metamaterial elements\textemdash thus, we use the radiation patterns of the antenna shown in Fig. \ref{Fig3_Patterns}(a) in the simulations. It is shown in Fig. \ref{Fig4_capacity}(a) that the sum capacity offered by the metasurface antenna is close to that by the Rayleigh channel. To evaluate the quality of the subchannels, we compute the ratio of the condition number, i.e., $\kappa_{MS}/\kappa_{Rayleigh}$. $\kappa_{MS}$ and $\kappa_{Rayleigh}$ represent the condition number of the channel matrix $\mathbf{H}$ in (\ref{SV_model}) and that of the Rayleigh channel matrix, respectively. Fig. \ref{Fig4_capacity}(b) shows the CDF of the ratio of the condition number for the angle spread $\sigma^{u}_{}=\sigma^{bs}_{}=20^{\circ}$, illustrating that the condition number of the MIMO channels provided by the proposed system is comparable to Rayleigh channel.

% proposed MIMO system using the metasurfaces provides the subchannels.
% as many uncorrelated subchannels as the Rayleigh channel can.
% In all cases the ICN reached is far below the i.i.d. Rayleigh cases. This means that, as might be expected, there is a loss of orthogonality with respect to the theoretical uncorrelated channels.

We also study the effects of the aperture size of the antenna and angular spread of the clusters on the performance of the proposed massive MIMO system. To alter the aperture size of the antenna, we sweep the number of metamaterial elements $N_{m}$ from 10 to 45, while fixing the number of metasurfaces to be $N_{bs}=20$ and the spacing between adjacent metamaterial elements to be $p_{m}=30$ mm. For each $N_{m}$, we simulate the metasurfaces using the CDM in Eq. (\ref{cdm_eq}) and compute their radiation patterns. We then use the computed radiation patterns in Eq. (\ref{SV_model}) to construct the channel matrix for each angle spread, varied from $5^{\circ}$ to $50^{\circ}$. For the analysis, we generated 10,000 channels for each value of angular spread. In the simulations, we fixed $\rho=10$ dB. 

% We then took the mean of the sum capacity, computed using Eq. (\ref{sum_capacity}), over the realized MIMO channels.
% metric

\begin{figure}[!t]
\centering
\includegraphics[width=3.20in]{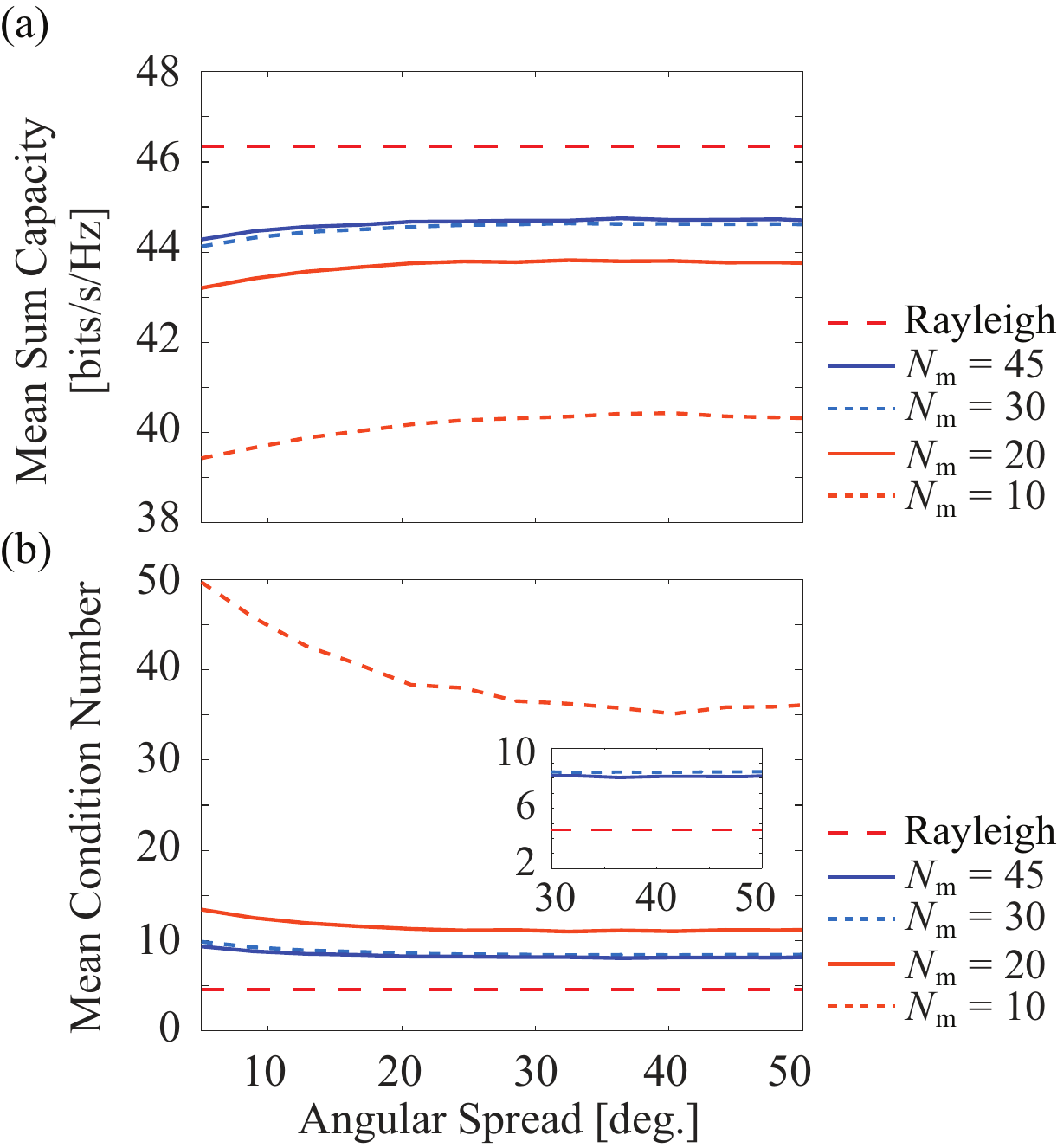}
\caption{Mean of (a) the sum capacity and, (b) the condition number of MIMO channel matrix. The aperture size of the antenna is varied by changing the number of metamaterial elements $N_{m}$, while fixing the number of metasurfaces $N_{bs}$ and the spacing $p_{m}$ between the elements. The angle spread $\sigma^{u}_{}=\sigma^{bs}_{}$ is also swept from $5^{\circ}$ to $60^{\circ}$.}
\label{Fig5_capacity_AS}
\end{figure}

In the analyses, we used the mean of the sum capacity over the realized MIMO channels as a metric to evaluate the performance of the proposed system. Fig. \ref{Fig5_capacity_AS}(a) shows the mean of the sum capacity of the proposed system for the swept angle spread (i.e., $\sigma^{u}_{}=\sigma^{bs}_{}$) and number of metamaterial elements $N_{m}$. As depicted in Fig. \ref{Fig5_capacity_AS}(a), the mean capacity increases as the number of the metamaterial elements increases, which can be attributed to the increased spatial resolution due to the increased aperture size. Note that the mean of the sum capacity is saturated for $N_{m} \geq 30$ and approaches $44.75$ bits/s/Hz, which is close to that offered by Rayleigh channel, $46.34$ bits/s/Hz. Also, for fixed $N_{m}$, the mean of the sum capacity is improved with increasing angular spread, as a result of the fixed spatial resolution of the metasurfaces.

Such observations in the mean of sum capacity in Fig. \ref{Fig5_capacity_AS}(a) can be demonstrated using the mean of the condition number of the realized channel matricies, as depicted in Fig. \ref{Fig5_capacity_AS}(b). As shown in Fig. \ref{Fig5_capacity_AS}(b), the increase in the aperture size (by increasing $N_{m}$) results in increased spatial resolution, leading to a reduced spatial correlation between subchannels. More specifically, the mean condition number of the proposed system reaches its minimum of $8.15$ for $\sigma^{u}=\sigma^{bs} \geq 30^{\circ}$. Note that the mean condition number of the Rayleigh channel is $4.53$. Also, it can be observed in Fig. \ref{Fig5_capacity_AS}(b) that the increase in the angle spread leads to the reduced correlation between subchannels for a fixed number of metamaterial elements $N_{m}$, which is the result of the fixed spatial resolution by the metasurfaces.

It should be noted that we have assumed static metamaterial elements throughout this work. Low-power driven tuning elements\textemdash such as diodes or liquid crystals\textemdash can be introduced to achieve dynamically reconfigurable radiation patterns. Such reconfigurable patterns can be used to reduce the correlation between subchannels and thereby optimize the performance of the uplink massive MIMO system. The analysis of the system performance using the dynamic metasurfaces is left as future work.

\section{Conclusion}
In this paper, we have proposed an uplink massive MIMO system using an array of rectangular waveguide-fed metasurfaces operating at $3.5$ GHz generating multiple fan beams. In the proposed massive MIMO system, the spatial properties of sub-6 GHz band MIMO channels, as well as the structural advantages of the metasurface antennas, are exploited to further reduce the RF complexity of the MIMO system. We also demonstrated the design of the metasurfaces using the holographic beamforming technique and the coupled dipole method. Using the designed antenna, we showed that a MIMO system using the metasurface antennas can offer a sum capacity approaching that of the Rayleigh channel. As the metasurface antennas are efficient, low-cost, and low-profile, the proposed massive MIMO system using the metasurfaces can find applications in various sub-6 GHz wireless networks requiring a large antenna system.

% reconfigurability
% intelligence
% precoding remains as future work.

% \section*{Acknowledgment}
% This work was supported by the National Science Foundation (NSF, Grant No.: 2030068).

\bibliographystyle{IEEEtran}
\bibliography{references.bib}

\end{document}